\documentclass[prc,superscriptaddress,unsortedaddress,twocolumn,showpacs,preprintnumbers,amsmath,amssymb,floatfix]{revtex4}
\usepackage[dvips]{graphicx}
\usepackage{amsmath}
\usepackage{amssymb}
\usepackage{times}
\usepackage{mathrsfs}
\usepackage{multirow}
\usepackage{bm}
\usepackage{wrapft}
\usepackage{color}

\def\ga{\mathrel{\mathpalette\fun >}}
\def\fun#1#2{\lower3.6pt\vbox{\baselineskip0pt\lineskip.9pt
\ialign{$\mathsurround=0pt#1\hfil##\hfil$\crcr#2\crcr\sim\crcr}}}

\newcommand{\beq}{\begin{equation}}
\newcommand{\eeq}{\end{equation}}
\newcommand{\bea}{\begin{eqnarray}}
\newcommand{\eea}{\end{eqnarray}}

\newcommand{\bfi}[1]{\mbox{\boldmath $#1$}}

\newcommand{\vK}{{\bfi K}}

\newcommand{\vs}{{\bfi s}}

\newcommand{\vrr}{{\bfi r}}
\newcommand{\vR}{{\bfi R}}

\begin{document}

\title{
Determination of the structure of $^{31}$Ne by
full-microscopic framework
}

\author{Kosho Minomo}
\affiliation{Department of Physics, Kyushu University, Fukuoka 812-8581, Japan}

\author{Takenori Sumi}
\affiliation{Department of Physics, Kyushu University, Fukuoka 812-8581, Japan}

\author{Masaaki Kimura}
\affiliation{Creative Research Institution (CRIS), Hokkaido University, Sapporo 001-0021, Japan}

\author{Kazuyuki Ogata}
\affiliation{Research Center of Nuclear Physics (RCNP), Osaka University, Ibaraki 567-0047, Japan}

\author{Yoshifumi R. Shimizu}
\affiliation{Department of Physics, Kyushu University, Fukuoka 812-8581, Japan}

\author{Masanobu Yahiro}
\affiliation{Department of Physics, Kyushu University, Fukuoka 812-8581, Japan}

\date{\today}

\begin{abstract}
We perform the first quantitative analysis of the reaction cross sections
of $^{28-32}$Ne by $^{12}$C
at 240~MeV/nucleon, using the double-folding model (DFM)
with the Melbourne $g$-matrix and
the deformed projectile density calculated by
the antisymmetrized molecular dynamics (AMD).
To describe the tail of the last neutron of $^{31}$Ne,
we adopt the resonating group method (RGM) combined with AMD.
The theoretical prediction excellently reproduce 
the measured cross sections of $^{28-32}$Ne with no adjustable parameters.
The ground state properties of $^{31}$Ne, i.e., strong
deformation and a halo structure with spin-parity $3/2_{}^-$,
are clarified.
\end{abstract}


\maketitle

{\it Introduction.}
Exotic properties of nuclei in the ``island of inversion''
fascinate many experimentalists and theoreticians.
The term ``island of inversion'' was first introduced
by Warburton {\it et al.}~\cite{Warburton} to specify
the region of unstable nuclei from $^{30}$Ne to $^{34}$Mg.
According to many experimental and theoretical studies,
it turned out that
the low excitation energies and the large $B(E2)$ values
of the first excited states of nuclei in the island
indicate strong
deformations~\cite{Fukunishi92,Mot95,Caurier,Utsuno,Iwas01,Yana03},
which eventually cause the {\it melt} of 
the neutron shell corresponding to the $N=20$ magic number
($N$: neutron number). 
To understand these features,
intruder configurations of the single-particle orbits
have been discussed with the shell model~\cite{Poves},
the generator coordinate method~\cite{Descouvemont},
and the Hartree-Fock (HF) and Hartree-Fock-Bogoliubov (HFB)
methods based on the Skyrme or Gogny
interaction~\cite{TFH97,YG04,RER02,RER03}.
In particular, the $^{31}$Ne nucleus is very interesting in view of
its intruder configurations and a halo structure due to strong deformations.
Recently, a systematic investigation employing
the antisymmetrized molecular dynamics (AMD) with the Gogny D1S interaction
has been performed for both even and odd $N$
nuclei in the ``island of inversion''~\cite{Kimura}.
AMD was shown to give rather large deformations and small separation
energy of the $^{31}$Ne as indicated by the preceding studies
mentioned above.

Very recently, experimental studies took large steps
toward exploring the ``island of inversion'', i.e.,
the one-neutron removal cross section $\sigma_{-n}$ of $^{31}$Ne
at 230~MeV/nucleon
was measured by Nakamura {\it et al}.~\cite{Nakamura} and
the interaction cross section $\sigma_{\rm I}^{}$ of $^{28-32}$Ne
by $^{12}$C at 240~MeV/nucleon was measured
by Takechi {\it et al}.~\cite{Takechi1,Takechi2}.
According to the analysis of the
$\sigma_{-n}^{}$~\cite{Nakamura,Horiuchi,Urata,ERT},
the measured large cross section suggested a neutron halo
structure of $^{31}$Ne with spin-parity $3/2_{}^-$.
This conjecture was confirmed by the analysis of the
$\sigma_{\rm I}^{}$~\cite{Takechi1,Takechi2}, in which neutron
configurations with lower partial waves were favored
to explain the $\sigma_{\rm I}^{}$ of $^{31}$Ne.
In Ref.~\cite{Takechi2}, it was shown that s-wave configuration
gave slightly better agreement with the data, which implied a possibility
of even more drastic change of the nuclear shell structure.
Note that $\sigma_{\rm I}$ differs from the total reaction cross section
$\sigma_{\rm R}$ by the inelastic cross section(s) due to
the excitation of the projectile. 
For unstable nuclei, however,
the two are almost identical, since very few bound excited
states exist for such nuclei. Thus, $\sigma_{\rm I}$,
as $\sigma_{\rm R}$, can be assumed to represent the {\it size} of the nucleus.

All aforementioned studies indicate the shell evolution of $^{31}$Ne
and its halo structure. Quantitative understanding of these properties
of $^{31}$Ne is, however, still under discussion. This is mainly due to
the fact that nuclear many-body wave functions obtained by the
high-precision structural models have never been directly applied
to reaction calculation.
As a first step, in Ref.~\cite{NeIsotopeDWS} we analyzed the
$\sigma_{\rm I}^{}$ of $^{28-32}$Ne~\cite{Takechi1,Takechi2}
with the microscopic double folding model (DFM) based on
the Melbourne $g$-matrix~\cite{Amos}.
We adopted the mean-field wave functions based on a deformed Woods-Saxon
potential, with the deformation parameter evaluated by AMD.
It was shown that the deformation of the Ne isotopes was indeed
important to reproduce the experimental data.
The agreement between the calculation and the data were, however, not
fully satisfactory.
The large difference in $\sigma_{\rm I}^{}$ between $^{30}$Ne, $^{31}$Ne,
and $^{32}$Ne could not be explained well in particular.

In this Letter,
we directly incorporate the AMD wave functions of $^{28-32}$Ne
in the DFM calculation and see how the structural properties
of these nuclei based on AMD are ``observed'' through
$\sigma_{\rm I}^{}$. For $^{31}$Ne, we further utilize
the resonating group method (RGM) to give a proper behavior
of the neutron wave function in the tail region.
This is the first microscopic calculation of
$\sigma_{\rm I}^{}$ for $^{28-32}$Ne with no
free parameters. The predicted values of $\sigma_{\rm I}^{}$
are validated by the comparison with the experimental
data~\cite{Takechi1,Takechi2}.
Through this study, we aim to determine the ground state
structure of $^{31}$Ne.

{\it Theoretical framework.}
We calculate the total reaction cross section $\sigma_{\rm R}^{}$
by DFM as in Ref.~\cite{NeIsotopeDWS}.
This model is accurate, when the projectile breakup is small. This 
condition is well satisfied for scattering analyses here, 
since the breakup cross section is quite small even for 
scattering of $^{31}$Ne with small neutron separation energy~\cite{ERT}. 
A microscopic optical potential $U$
between a projectile (P) and a target (T) is
constructed by folding the effective nucleon-nucleon ($NN$) interaction with
the projectile and target densities, $\rho_{\rm P}^{}$ and $\rho_{\rm T}^{}$,
respectively. 
The direct ($U_{\rm DR}$) and exchange ($U_{\rm EX}$)
parts of the folding potential are obtained
by~\cite{DFM-standard-form,DFM-standard-form-2}
\bea
\label{eq:UD}
U_{\rm DR}^{}(\vR)&=&\int
 \rho_{\rm P}^{}(\vrr_{\rm P}) \rho_{\rm T}^{}(\vrr_{\rm T})
  v_{\rm DR}^{}(\rho,\vs) d\vrr_{\rm P}^{} d\vrr_{\rm T}^{}, \\
\label{eq:UEX}
U_{\rm EX}^{}(\vR)&=&\int
 \rho_{\rm P}^{}(\vrr_{\rm P}^{},\vrr_{\rm P}^{}+\vs)
  \rho_{\rm T}^{}(\vrr_{\rm T}^{},\vrr_{\rm T}^{}-\vs) \nonumber \\
   &&~~\times v_{\rm EX}^{}(\rho,\vs) \exp{[i\vK(\vR) \! \cdot \! \vs/M]}
    d\vrr_{\rm P}^{} d\vrr_{\rm T}^{},~~~~
\eea
where $\vs=\vrr_{\rm P}-\vrr_{\rm T}+\vR$ for a position vector $\vR$ of
the center-of-mass of P from that of T.
The original form of $U_{\rm EX}$ is a non-local function of $\vR$,
but it has been localized in Eq.~\eqref{eq:UEX} with the local semi-classical
approximation~\cite{Brieva-Rook}, where $\hbar \vK(\vR)$ is the local momentum
of the scattering considered and
$M=A_{\rm P}A_{\rm T}/(A_{\rm P}+A_{\rm T})$ for the mass number
$A_{\rm P}$ ($A_{\rm T}$) of P (T).
The validity of this localization is shown in Ref.~\cite{Minomo:2009ds}
for nucleon-nucleus scattering; note that this is also the case with
nucleus-nucleus scattering.
In Eqs.~\eqref{eq:UD} and~\eqref{eq:UEX}
the direct (exchange) component of the effective $NN$ interaction,
$v_{\rm DR}$ ($v_{\rm EX}$),
is assumed to depend on the local density
at the midpoint of the interacting nucleon pair. We adopt
the Melbourne $g$-matrix as an effective $NN$ interaction in
nuclear medium.

As for $\rho_{\rm T}^{}$, we use the phenomenological $^{12}$C-density
deduced from the electron scattering~\cite{C12-density}, with
unfolded the finite-size effect of the proton charge following the
standard manner~\cite{Singhal}.

The projectile densities $\rho_{\rm P}$ of $^{28-32}$Ne are calculated 
from the AMD wave functions that successfully describe the low-lying 
spectrum of Ne isotopes \cite{Kimura}. The reader is directed to 
them for the details of the AMD wave function 
and calculation of $\rho_{\rm P}$. 
To investigate one neutron-halo nature, we have performed more 
sophisticated calculation for $^{31}$Ne, that is called AMD-RGM below. 
Employing the coupled-channels RGM type wave function, 
\begin{align}
 \Psi(^{31}{\rm Ne}; 3/2^-_1) = 
  \sum_{nJ\pi}{\cal A}
  \left\{
   \chi_{nl}^{}(r) Y_{lm}^{}(\hat{\vrr})
    \Psi(^{30}{\rm Ne}; J^\pi_n)\phi_{n}^{}
  \right\},
\end{align}
the relative wave function $\chi_{nj}^{}$ between the last neutron 
and the core ($^{30}$Ne) is calculated by solving the RGM equation. 
Here the wave functions of $^{30}$Ne are those of AMD obtained in 
Ref \cite{Kimura} and includes the many excited states with positive- and
negative-parity below 10 MeV in excitation energy. 
Therefore, note that the weak-binding feature and the possible 
core excitation associated with the strong deformation are properly
treated in the AMD-RGM calculation.

If one or both of the densities $\rho_{\rm P}^{}$ and
$\rho_{\rm T}^{}$ are non-spherical,
the microscopic potential $U$ is not spherical.
It follows from Ref.~\cite{Satchler-1979,NeIsotopeDWS}, however,
that it is sufficient to use an angular-averaged density in DFM,
i.e., Eqs.~\eqref{eq:UD} and~\eqref{eq:UEX}, in the present case.
Nevertheless, deformation effects of the wave functions are
taken into account as shown in Ref.~\cite{NeIsotopeDWS}.

\begin{figure}[htbp]
\begin{center}
 \includegraphics[width=0.35\textwidth,clip]{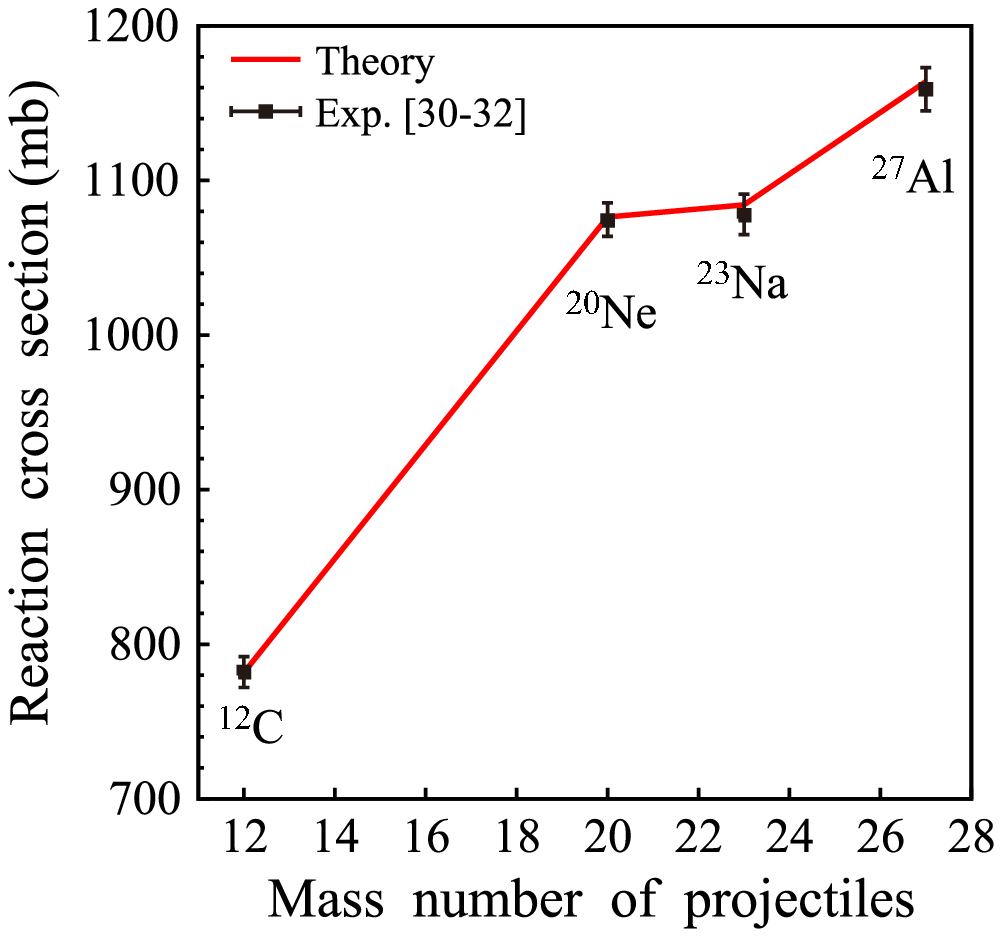}
 \caption{(Color online)
Reaction cross sections for scattering of stable nuclei from
$^{12}$C at 250~MeV/nucleon.
The solid line represents the results of
the present double folding model calculation.
The experimental data are taken from Ref.~\cite{expC12C12,20Ne,23Na}.
}
 \label{fig1}
\end{center}
\end{figure}

{\it Results and Discussions.}
We show in Fig.~\ref{fig1} the result of
$\sigma_{\rm R}^{}$ for $^{12}$C, $^{20}$Ne, $^{23}$Na, and $^{27}$Al
by a $^{12}$C target at around 250~MeV/nucleon, compared with the experimental
data~\cite{expC12C12,20Ne,23Na}. 
For $^{20}$Ne and $^{23}$Na, the original experimental data were 
measured at around 950~MeV/nucleon~\cite{20Ne,23Na} but the present ones are 
corrected in 240~MeV/nucleon by Glauber calculation~\cite{expC12C12}. 
The projectile densities are phenomenological ones
obtained by electron scattering~\cite{C12-density}.
The theoretical results of $\sigma_{\rm R}^{}$ shown in this Letter
are reduced by 1.8\% as in Ref.~\cite{NeIsotopeDWS}. 
This fine tuning has been done to reproduce the mean value of
the $\sigma_{\rm R}^{}$ of the $^{12}$C-$^{12}$C scattering measured
at 250.8~MeV/nucleon~\cite{expC12C12}. Thus, the present calculation
contains no free parameters except for the $^{12}$C-$^{12}$C
scattering. Figure~\ref{fig1} shows the high accuracy of the
$\sigma_{\rm R}^{}$ predicted by the present DFM calculation for
the $^{20}$Ne, $^{23}$Na, and $^{27}$Al projectiles.

\begin{table}[htbp]
\caption{
The spin-parity and deformation parameter $\beta$ and $\gamma$
of the ground states of Ne isotopes calculated by AMD. }
\begin{center}
\begin{tabular}{ccccccc} \hline
 nuclide  & ~ & $^{28}$Ne & $^{29}$Ne & $^{30}$Ne & $^{31}$Ne
 & $^{32}$Ne \\ \hline
 $J^\pi$  & ~ &$0^+$ & $1/2^+$ & $0^+$ & $3/2^-$ & $0^+$ \\
 $\beta$  & ~ & 0.28     & 0.43      & 0.39     & 0.41
 & 0.33      \\ 
 $\gamma$ & ~ & 60$^\circ$ & 0$^\circ$ & 0$^\circ$ & 0$^\circ$
 & 0$^\circ$  \\ \hline
\end{tabular}
\label{tab1}
\end{center}
\end{table}
%
\begin{figure}[htbp]
\begin{center}
 \includegraphics[width=0.35\textwidth,clip]{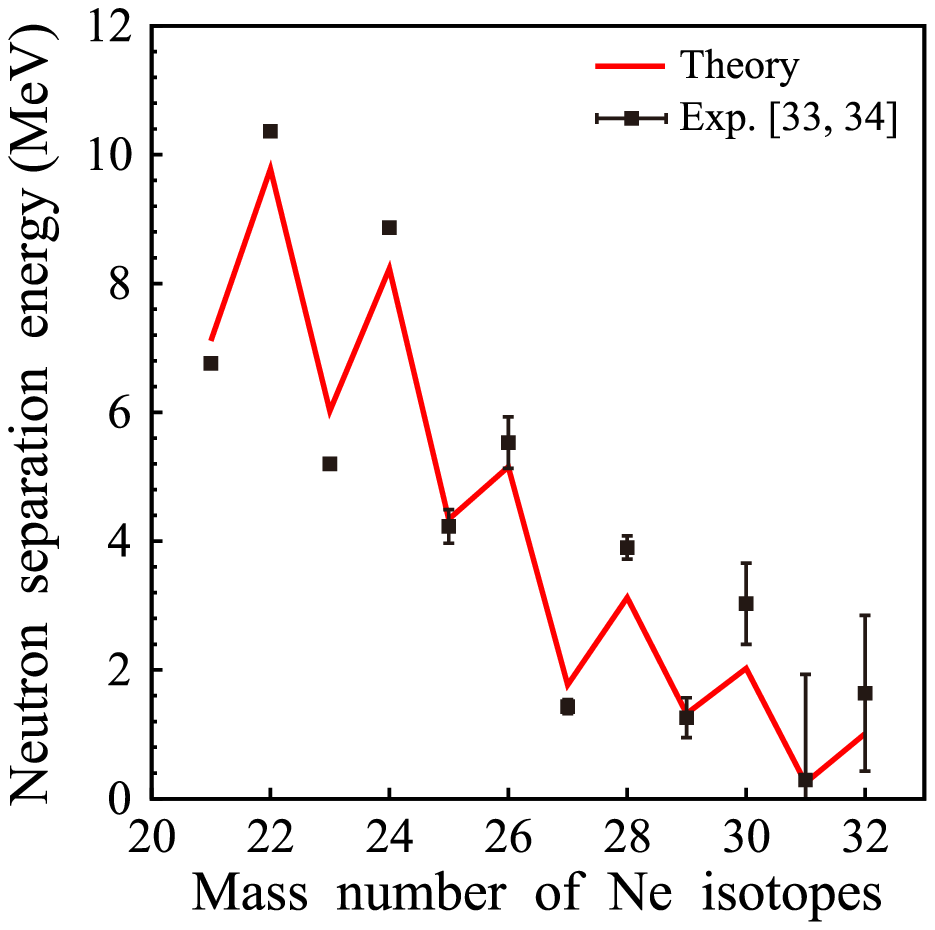}
 \caption{(Color online)
One neutron separation energy $S_n$ of the Ne isotopes.
The experimental data are taken from Ref.~\cite{Audi, Sn}. 
 }
 \label{fig2}
\end{center}
\end{figure}
The structural properties of $^{28-32}$Ne obtained by
AMD are shown in
Table~\ref{tab1} (spin-parity and deformation parameter 
$\beta$ and $\gamma$ defined by the Hill-Wheeler coordinate) and
Fig.~\ref{fig2} (one neutron separation energies $S_n$).
In the latter, experimental data~\cite{Audi,Sn} are also shown.
One clearly sees that $^{28-32}$Ne are strongly deformed
and the odd-even staggering of $S_n$ measured is reproduced
very well. It is thus expected that the AMD wave functions
of Ne isotopes are highly reliable. However, for $^{31}$Ne,
which has very small $S_n$ ($\sim 250$~keV), the tail of the
wave function of the last neutron may not be described properly, 
since AMD uses a one-range Gaussian wave function for the motion of
each nucleon. 

\begin{figure}[htbp]
\begin{center}
 \includegraphics[width=0.35\textwidth,clip]{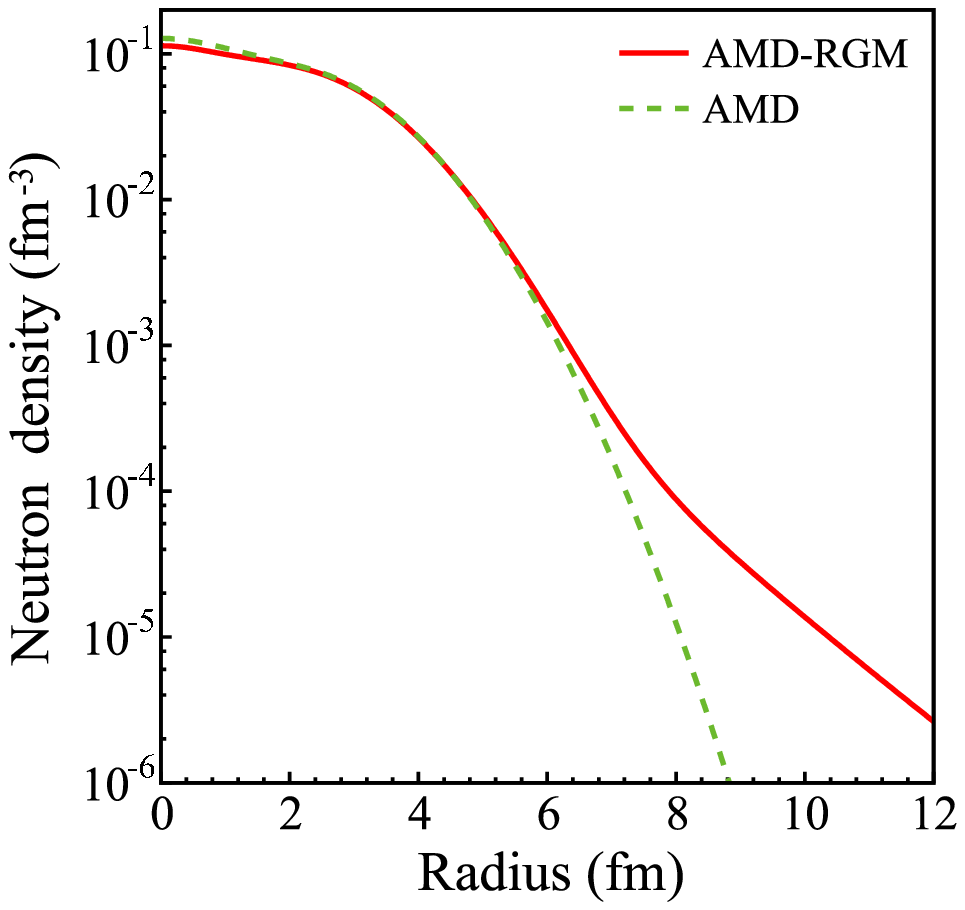}
 \caption{(Color online)
The neutron one-body densities of $^{31}$Ne.
The solid and dashed lines represent the results of $^{31}$Ne
calculated by AMD-RGM and AMD, respectively. }
 \label{fig3}
\end{center}
\end{figure}

This possible shortcomings can be overcome by using RGM, which
generates a proper asymptotics of the last neutron. 
The neutron one-body density $\rho_n (r)$ of
$^{31}$Ne thus obtained is shown in Fig.~\ref{fig3} by the solid line.
The results for $^{31}$Ne (dashed line) 
without RGM are also shown for comparison.
The solid line has a long tail, whereas the dashed line rapidly
falls off at $r \ga 6$~fm. 
The root mean square radius of the density obtained by
AMD-RGM (AMD) is 3.617~fm (3.490~fm). Although the difference
between the two values is not so large, the density in
the tail region, i.e., $r \ga 6$~fm, has a significant
contribution to $\sigma_{\rm R}^{}$ as shown below.
Note that AMD-RGM gives $S_n =450$~keV, which is slightly
larger than 250~keV obtained by AMD but still consistent
with the measured value, $S_n^{}=0.29 \pm 1.64$~MeV~\cite{Sn}.

\begin{table}[htbp]
\begin{center}
\caption{Configurations of the ground state of $^{31}$Ne
obtained by AMD-RGM and AMD. }
\begin{tabular}{ccc}
\hline \hline
~~~~~Spherical basis~~~~~ & \multicolumn{2}{c}{Amplitude} \\ \cline{2-3}
 {} & ~~~AMD-RGM~~~ & ~~~~AMD~~~~ \\ \hline
$^{30}$Ne($0_{}^+$) $\otimes$ $1p_{3/2}^{}$ & 56 \% & 37 \% \\
$^{30}$Ne($2_{}^+$) $\otimes$ $1p_{3/2}^{}$ & 24 \% & 41 \% \\
$^{30}$Ne($2_{}^+$) $\otimes$ $0f_{7/2}^{}$ & ~9 \% & 12 \% \\
$^{30}$Ne($1_{}^-$) $\otimes$ $1s_{1/2}^{}$ & ~5 \% & ~5 \% \\
other components & ~6 \% & ~5 \% \\ \hline \hline
\end{tabular}
\label{tab2}
\end{center}
\end{table}

The AMD-RGM and AMD wave functions of $^{31}$Ne can be decomposed
in terms of the spherical basis components as shown in Table~\ref{tab2}.
Compared to the AMD, the amount of $2^+$ states reduces in AMD-RGM. 
This is due to the weak-coupling between $^{30}$Ne and valence neutron. 
In the AMD-RGM model, the main component of the last neutron is $1p_{3/2}^{}$
coupled with the $0^+$ ground state of $^{30}$Ne. 
The long range tail of $\rho_n (r)$, i.e., the halo structure,
shown in Fig.~\ref{fig3} is due to this configuration of $^{31}$Ne.
Thus, reliable nuclear wave functions of Ne isotopes, including
a p-wave halo nucleus $^{31}$Ne, are prepared. 

\begin{figure}[htbp]
\begin{center}
 \includegraphics[width=0.35\textwidth,clip]{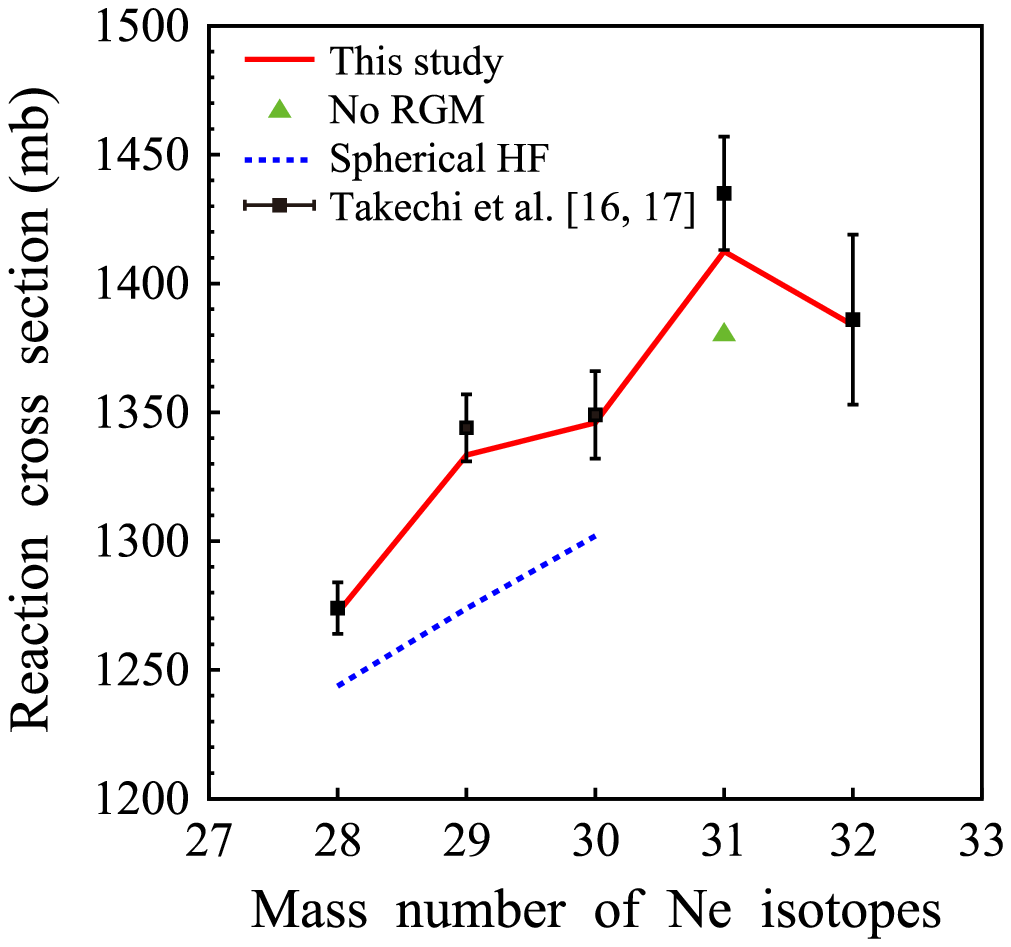}
 \caption{(Color online)
Reaction cross sections of Ne isotopes by
$^{12}$C at 240~MeV/nucleon.
The solid line represents the result of the present study.
The result obtained by HF calculation is shown by the
dashed line.
The triangle represents the result for $^{31}$Ne without RGM prescription.
The experimental data are taken from Refs.~\cite{Takechi1,Takechi2}.
}
\label{fig4}
\end{center}
\end{figure}
In Fig.~\ref{fig4} we show by the solid line the prediction
of $\sigma_{\rm R}^{}$ for $^{28-32}$Ne + $^{12}$C scattering
at 240~MeV/nucleon. 
The agreement of the present result with the experimental data is excellent.
If we adopt spherical HF wave functions for Ne isotopes,
the dashed line is obtained, which significantly undershoots
the data. More seriously, no bound-state solution is found
for $^{31,32}$Ne. The triangle shows the result for $^{31}$Ne
calculated with AMD wave functions. 
Clearly, the result
cannot explain the very large experimental value.
We conclude from these findings that i) $^{28-32}$Ne are
strongly deformed as discussed in Ref.~\cite{NeIsotopeDWS}
and ii) $^{31}$Ne has a halo structure due to the last neutron
in the $1p_{3/2}^{}$ orbit.
Very recently,
in Ref.~\cite{Hagino} the paring anti-halo effect was discussed
for odd-even staggering of $\sigma_{\rm R}$ for $^{30-32}$Ne.
It will be interesting to consider the pairing effects in the
present framework. To see how the strong deformations indicated
by AMD affect the result in Ref.~\cite{Hagino} will also be
very important. 

{\it Summary.}
We have performed a microscopic calculation of
the reaction cross sections for neutron-rich Ne-isotopes systematically.
The double-folding model (DFM) with the Melbourne $g$-matrix
and the antisymmetrized molecular dynamics (AMD) wave functions were used.
AMD is a powerful tool that describes strongly deformed nuclei.
For a loosely-bound nucleus $^{31}$Ne, the resonating group method
(RGM) was adopted to generate a proper behavior of the
wave function of the last neutron.
The present framework reproduced the experimental data very well
with no free adjustable parameters. Thus, the AMD wave functions of the
Ne isotopes (AMD-RGM for $^{31}$Ne) were clearly validated.
We concluded that neutron-rich Ne-isotopes are strongly
deformed and $^{31}$Ne has a halo structure with
spin-parity $3/2_{}^-$.
In the near future, we will apply the present framework to
investigate properties of Mg-isotopes, for further exploration of
the ``island of inversion''.

\section*{Acknowledgements}

The authors thank M.~Takechi for providing the values of
the experimental data and H. Sakurai and M. Fukuda for useful discussions.
This work is supported in part by Grant-in-Aid for Scientific Research~(C)
No.~22540285 and 22740169 from Japan Society for the Promotion of Science.
The numerical calculations of this work were performed
on the computing system in Research Institute
for Information Technology of Kyushu University.



\begin{thebibliography}{00}

\bibitem{Warburton}
E.~K.~Warburton, J.~A.~Becker, and B.~A.~Brown,
Phys.\ Rev.\ C {\bf 41}, 1147 (1990).

\bibitem{Fukunishi92}
N. Fukunishi, T. Otsuka, and T. Sebe,
Phys.\ Lett.\ B {\bf 296}, 279 (1992).

\bibitem{Mot95}
T. Motobayashi {\it et al.},
Phys.\ Lett.\ B {\bf 346}, 9 (1995).

\bibitem{Caurier}
E.~Caurier, F. Nowacki, A. Poves, J. Retamosa,
Phys.\ Rev.\ C {\bf 58}, 2033 (1998).

\bibitem{Utsuno}
Y.~Utsuno, T. Otsuka, T. Mizusaki, M. Honma,
Phys.\ Rev.\ C {\bf 60}, 054315 (1999).

\bibitem{Iwas01}
H. Iwasaki {\it et al.},
Phys.\ Lett.\ B {\bf 522}, 227 (2001).

\bibitem{Yana03}
Y. Yanagisawa {\it et al.},
Phys.\ Lett.\ B {\bf 566}, 84 (2003).

\bibitem{Poves}
A. Poves and J. Retamosa,
Nucl. Phys. A {\bf 571}, 221 (1994).

\bibitem{Descouvemont}
P. Descouvemont,
Nucl. Phys. A {\bf 655}, 440 (1999).

\bibitem{TFH97}
J. Terasaki, H. Flocard, P.-H. Heenen, and P. Bonche,
Nucl.\ Phys.\ A {\bf 621}, 706 (1997).

\bibitem{RER02}
R.~Rodr\'iguez-Guzm\'an, J.L. Egido, and L.M. Robledo,
Nucl.\ Phys.\ A {\bf 709}, 201 (2002).

\bibitem{YG04}
M. Yamagami and Nguyen Van Giai,
Phys.\ Rev.\ C {\bf 69}, 034301 (2004).

\bibitem{RER03}
R.R.~Rodr\'iguez-Guzm\'an, J.L. Egido, and L.M. Robledo,
Eur.\ Phys.\ J.\ A {\bf 17}, 37 (2003).

\bibitem{Kimura}
M. Kimura and H. Horiuchi, Prog. Theor. Phys. 111, 841 (2004);
M. Kimura, Phys. Rev. C 75, 041302 (2007).
M. Kimura, arXiv:1105.3281 (2011) [nucl-th].

\bibitem{Nakamura}
T. Nakamura, \textit{et al}.,
Phys. Rev. Lett. {\bf 103}, 262501 (2009).

\bibitem{Takechi1}
M. Takechi {\it et al.}, Nucl. Phys. {\bf A834}, 412c (2010).

\bibitem{Takechi2}
M. Takechi {\it et al.}, Mod. Phys. Lett. A {\bf 25}, 1878 (2010).

\bibitem{Horiuchi}
W. Horiuchi, Y. Suzuki, P. Capel, and D. Baye,
Phys. Rev. C {\bf 81}, 024606 (2010).

\bibitem{Urata}
Y. Urata, K. Hagino, and H. Sagawa,
Phys. Rev. C {\bf 83}, 041303(R) (2011).

\bibitem{ERT}
M. Yahiro, K. Ogata, and K. Minomo,
Prog. Theor. Phys. {\bf 126}, 167 (2011). 

\bibitem{NeIsotopeDWS}
K. Minomo, T. Sumi, M. Kimura, K. Ogata, Y. R. Shimizu, and M. Yahiro,
Phys. Rev. C {\bf 84}, 034602 (2011).

\bibitem{Amos}
K. Amos, P. J. Dortmans, H. V. von Geramb, S. Karataglidis,
and J. Raynal, Adv. Nucl. Phys. {\bf 25}, 275 (2000).

\bibitem{DFM-standard-form}
B. Sinha, Phys. Rep. {\bf 20}, 1 (1975). \\
B. Sinha and S. A. Moszkowski, Phys. Lett. B{\bf 81}, 289 (1979).

\bibitem{DFM-standard-form-2}
T. Furumoto, Y. Sakuragi, and Y. Yamamoto, Phys. Rev. C{\bf 82},
044612 (2010).

\bibitem{Brieva-Rook}
F.A. Brieva and J.R. Rook, Nucl. Phys. A{\bf 291}, 299 (1977);
ibid. 291, 317 (1977); ibid. 297, 206 (1978).

\bibitem{Minomo:2009ds}
K.~Minomo, K.~Ogata, M.~Kohno, Y.~R.~Shimizu, and M.~Yahiro,
J.\ Phys.\ G {\bf 37}, 085011 (2010)
[arXiv:0911.1184 [nucl-th]].

 \bibitem{C12-density} H. de Vries, C. W. de Jager, and C. de Vries,
 At. Data Nucl. Data Tables \textbf{36}, 495 (1987).

\bibitem{Singhal}
R. P. Singhal \textit{et al}., Nucl. Instr. and Meth. 148, 113(1978).

\bibitem{Satchler-1979}
G. R. Satchler, Phys. Rep. {\bf 55}, 183-254 (1979).

 \bibitem{expC12C12}
  M. Takechi, \textit{et al}.,
  Phys.\ Rev.\ C \textbf{79}, 061601(R) (2009).

\bibitem{20Ne}
L. Chulkov \textit{et al}.,
Nucl. Phys. A {\bf 603}, 219 (1996).

\bibitem{23Na}
T. Suzuki \textit{et al}.,
Phys. Rev. Lett. {\bf 75}, 3241 (1995).

\bibitem{Sn}
B. Jurado, \textit{et al}., Phys. Lett. B \textbf{649}, 43 (2007).

\bibitem{Audi}
G. Audi, A. H. Wapstra, and C. Thibault, 
Nucl. Phys. A {\bf 729}, 337 (2003). 

\bibitem{Hagino}
K. Hagino and H. Sagawa,
Phys. Rev. C {\bf 84}, 011303(R) (2011).



\end{thebibliography}
\end{document}